\documentclass[11pt]{article}

\catcode`\@=11
\def\@citex[#1]#2{%
\if@filesw \immediate \write \@auxout {\string \citation {#2}}\fi
\@tempcntb\m@ne \let\@h@ld\relax \def\@citea{}%
\@cite{%
  \@for \@citeb:=#2\do {%
    \@ifundefined {b@\@citeb}%
      {\@h@ld\@citea\@tempcntb\m@ne{\bf ?}%
      \@warning {Citation `\@citeb ' on page \thepage \space undefined}}%
      {\@tempcnta\@tempcntb \advance\@tempcnta\@ne%
      \@tempcntb\number\csname b@\@citeb \endcsname \relax%
      \ifnum\@tempcnta=\@tempcntb 
        \ifx\@h@ld\relax%
          \edef \@h@ld{\@citea\csname b@\@citeb\endcsname}%
        \else%
          \edef\@h@ld{\ifmmode{-}\else--\fi\csname b@\@citeb\endcsname}%
        \fi%
      \else
        \@h@ld\@citea\csname b@\@citeb \endcsname%
        \let\@h@ld\relax%
      \fi}%
    \def\@citea{,\penalty\@highpenalty\,}%
  }\@h@ld
}{#1}}

\def\@citeb#1#2{{[#1]\if@tempswa , #2\fi}}
\def\@citeu#1#2{{$^{#1}$\if@tempswa , #2\fi }}
\def\@citep#1#2{{#1\if@tempswa , #2\fi}}

\def\bcites{         
        \catcode`\@=11
        \let\@cite=\@citeb
        \catcode`\@=12
}

\def\upcites{         
        \catcode`\@=11
        \let\@cite=\@citeu
        \catcode`\@=12
}

\def\plaincites{      
        \catcode`\@=11
        \let\@cite=\@citep
        \catcode`\@=12
}

\newcount\hour
\newcount\minute
\newtoks\amorpm
\hour=\time\divide\hour by 60
\minute=\time{\multiply\hour by 60 \global\advance\minute by-\hour}
\edef\standardtime{{\ifnum\hour<12 \global\amorpm={am}%
        \else\global\amorpm={pm}\advance\hour by-12 \fi
        \ifnum\hour=0 \hour=12 \fi
        \number\hour:\ifnum\minute<10 0\fi\number\minute\the\amorpm}}
\edef\militarytime{\number\hour:\ifnum\minute<10 0\fi\number\minute}

\def\draftlabel#1{{\@bsphack\if@filesw {\let\thepage\relax
   \xdef\@gtempa{\write\@auxout{\string
      \newlabel{#1}{{\@currentlabel}{\thepage}}}}}\@gtempa
   \if@nobreak \ifvmode\nobreak\fi\fi\fi\@esphack}
        \gdef\@eqnlabel{#1}}
\def\@eqnlabel{}
\def\@vacuum{}
\def\marginnote#1{}
\def\draftmarginnote#1{\marginpar{\raggedright\scriptsize\tt#1}}
\def\draft{
        \pagestyle{plain}
        \overfullrule=2pt
        \oddsidemargin -.5truein
        \def\@oddhead{\sl \phantom{\today\quad\militarytime} \hfil
        \smash{\Large\sl DRAFT} \hfil \today\quad\militarytime}
        \let\@evenhead\@oddhead
        \let\label=\draftlabel
        \let\marginnote=\draftmarginnote
        \def\ps@empty{\let\@mkboth\@gobbletwo
        \def\@oddfoot{\hfil \smash{\Large\sl DRAFT} \hfil}
        \let\@evenfoot\@oddhead}
        \def\@eqnnum{(\theequation)\rlap{\kern\marginparsep\tt\@eqnlabel}%
        \global\let\@eqnlabel\@vacuum}  }

\def\blackfonts{
        \font\blackboard=msbm10 scaled\magstep1
        \font\blackboards=msbm8
        \font\blackboardss=msbm6
}

\def\prep{         
        \catcode`\@=11
        \input art10.sty
        \catcode`\@=12
        
        \let\small\null
        \def\blackfonts{
                \font\blackboard=msbm10
                \font\blackboards=msbm7
                \font\blackboardss=msbm5
        }
        \let\sl\it
        \twocolumn
        \sloppy
        \voffset=-2.54truecm
        \hoffset=-2.54truecm
        \flushbottom
        \parindent 1em
        \leftmargini 2em
        \leftmarginv .5em
        \leftmarginvi .5em
        \marginparwidth 48pt
        \marginparsep 10pt
        \setlength{\columnsep}{2truecm}
        \setlength{\textwidth}{25.4truecm}
        \setlength{\textheight}{17truecm}
        \baselineskip=16pt
        \oddsidemargin .18truein
        \evensidemargin .17truein
}

\def\eqalign#1{\null\,\vcenter{\openup\jot\m@th
  \ialign{\strut\hfil$\displaystyle{##}$&$\displaystyle{{}##}$\hfil
      \crcr#1\crcr}}\,}
\def\eqalignno#1{\displ@y \tabskip\centering
  \halign to\displaywidth{\hfil$\@lign\displaystyle{##}$\tabskip\z@skip
    &$\@lign\displaystyle{{}##}$\hfil\tabskip\centering
    &\llap{$\@lign##$}\tabskip\z@skip\crcr
    #1\crcr}}

\def\section{\@startsection {section}{1}{\z@}{3.ex plus 1ex minus
 .2ex}{2.ex plus .2ex}{\large\bf}}
\def\subsection{\@startsection{subsection}{2}{\z@}{2.75ex plus 1ex minus
 .2ex}{1.5ex plus .2ex}{\bf}}        

\def\appendix{{\newpage\section*{Appendix}}\let\appendix\section%
        {\setcounter{section}{0}
        \gdef\thesection{\Alph{section}}}\section}

\def\abstract{\if@twocolumn
\section*{Abstract}
\else 
\begin{center}
{\bf Abstract\vspace{-.5em}\vspace{0pt}}
\end{center}
\quotation
\fi}

\catcode`\@=12

\def\d{\partial}

\def\sqr#1#2{{\vcenter{\vbox{\hrule height.#2pt\hbox{\vrule width.#2pt 
height#1pt \kern#1pt \vrule width.#2pt}\hrule height.#2pt}}}}
 
\def\fii{\varphi}

\def\=d{\,{\buildrel\rm def\over =}\,}

\def\A{\hbox{\bbf A}}

\def\i3p{\p32\int d^3p}

\def\As{A\hbox to 1pt{\hss /}}
\def\np4{\int d^4p_1\cdots d^4p_{n-1}\, }

\def\nx4{\int d^4x_1\ldots d^4x_n\, }

\def\kon#1#2{\vbox{\halign{##&&##\cr
\lower4pt\hbox{$\scriptscriptstyle\vert$}\hrulefill &
\hrulefill\lower4pt\hbox{$\scriptscriptstyle\vert$}\cr $#1$&
$#2$\cr}}}

\def\konv#1#2#3{\hbox{\vrule height12pt depth-1pt}
\vbox{\hrule height12pt width#1cm depth-11.6pt}
\hbox{\vrule height6.5pt depth-0.5pt}
\vbox{\hrule height11pt width#2cm depth-10.6pt\kern5pt
      \hrule height6.5pt width#2cm depth-6.1pt}
\hbox{\vrule height12pt depth-1pt}
\vbox{\hrule height6.5pt width#3cm depth-6.1pt}
\hbox{\vrule height6.5pt depth-0.5pt}}
\def\konu#1#2#3{\hbox{\vrule height12pt depth-1pt}
\vbox{\hrule height1pt width#1cm depth-0.6pt}
\hbox{\vrule height12pt depth-6.5pt}
\vbox{\hrule height6pt width#2cm depth-5.6pt\kern5pt
      \hrule height1pt width#2cm depth-0.6pt}
\hbox{\vrule height12pt depth-6.5pt}
\vbox{\hrule height1pt width#3cm depth-0.6pt}
\hbox{\vrule height12pt depth-1pt}}

\def\konw#1#2#3{\hbox{\vrule height12pt depth-1pt}
\vbox{\hrule height12pt width#1cm depth-11.6pt}
\hbox{\vrule height6.5pt depth-0.5pt}
\vbox{\hrule height12pt width#2cm depth-11.6pt \kern5pt
      \hrule height6.5pt width#2cm depth-6.1pt}
\hbox{\vrule height6.5pt depth-0.5pt}
\vbox{\hrule height12pt width#3cm depth-11.6pt}
\hbox{\vrule height12pt depth-1pt}}

\def\i{{\rm int}}

\def\e{{\rm ext}}

\def\r{{\rm ret}}
\def\a{{\rm av}}

\def\m3{{\mu_1\mu_2\mu_3}}

\def\p{{(+)}}

\def\be{\begin{equation}}       \def\eq{\begin{equation}}
\def\ee{\end{equation}}         \def\eqe{\end{equation}}

\def\bea{\begin{eqnarray}}      \def\eqa{\begin{eqnarray}}
\def\ena{\end{eqnarray}}        \def\eea{\end{eqnarray}}
                                \def\eqae{\end{eqnarray}}

\def\ba{\begin{array}}
\def\ea{\end{array}}
\def\unit{1 \hskip-.3em \raise2pt\hbox{$ \scriptstyle |$ } }

\def\a{\alpha}
\def\b{\beta}
 
\def\d{\delta}
\def\e{\epsilon}           

\def\i{\iota}
\def\j{\psi}
\def\l{\lambda}
\def\m{\mu}
\def\n{\nu}
  
\def\p{\pi}                
\def\r{\rho}                                     
\def\s{\sigma}                                   
\def\t{\tau}

\def\D{\Delta}

\def\bop#1{\setbox0=\hbox{$#1M$}\mkern1.5mu
        \vbox{\hrule height0pt depth.04\ht0
        \hbox{\vrule width.04\ht0 height.9\ht0 \kern.9\ht0
        \vrule width.04\ht0}\hrule height.04\ht0}\mkern1.5mu}

\def\>{\rangle} 

\def\<{\langle} 
\def\Dsl{D \hskip-.6em \raise1pt\hbox{$ / $ } }

\def\sl#1{\rlap{\hbox{$\mskip 1 mu /$}}#1}
\def\leftrightarrowfill{$\mathsurround=0pt \mathord\leftarrow \mkern-6mu
       \cleaders\hbox{$\mkern-2mu \mathord- \mkern-2mu$}\hfill
       \mkern-6mu \mathord\rightarrow$}
\def\dvec#1{\vbox{\ialign{##\crcr
       \leftrightarrowfill\crcr\noalign{\kern-1pt\nointerlineskip}
       $\hfil\displaystyle{#1}\hfil$\crcr}}}          
\def\hook#1{{\vrule height#1pt width0.4pt depth0pt}}
\def\leftrighthookfill#1{$\mathsurround=0pt \mathord\hook#1
       \hrulefill\mathord\hook#1$}
\def\underhook#1{\vtop{\ialign{##\crcr                 
       $\hfil\displaystyle{#1}\hfil$\crcr
       \noalign{\kern-1pt\nointerlineskip\vskip2pt}
       \leftrighthookfill5\crcr}}}
\def\smallunderhook#1{\vtop{\ialign{##\crcr      
       $\hfil\scriptstyle{#1}\hfil$\crcr
       \noalign{\kern-1pt\nointerlineskip\vskip2pt}
       \leftrighthookfill3\crcr}}}

\def\sfrac#1#2{{\vphantom1\smash{\lower.5ex\hbox{\small$#1$}}\over
       \vphantom1\smash{\raise.4ex\hbox{\small$#2$}}}} 
\def\bfrac#1#2{{\vphantom1\smash{\lower.5ex\hbox{$#1$}}\over
       \vphantom1\smash{\raise.3ex\hbox{$#2$}}}}      
\def\afrac#1#2{{\vphantom1\smash{\lower.5ex\hbox{$#1$}}\over#2}}  
\def\on#1#2{{\buildrel{\mkern2.5mu#1\mkern-2.5mu}\over{#2}}}
\def\ddt#1{\on{\hbox{\LARGE .\kern-2pt.}}#1}             
\def\tdt#1{\on{\hbox{\LARGE .\kern-2pt.\kern-2pt.}}#1}   

\def\boxes#1{
       \newcount\num
       \num=1
       \newdimen\downsy
       \downsy=-1.5ex
       \mskip-2.8mu
       \bo
       \loop
       \ifnum\num<#1
       \llap{\raise\num\downsy\hbox{$\bo$}}
       \advance\num by1
       \repeat}
\def\boxup#1#2{\newcount\numup
       \numup=#1
       \advance\numup by-1
       \newdimen\upsy
       \upsy=.75ex
       \mskip2.8mu
       \raise\numup\upsy\hbox{$#2$}}

\newskip\humongous \humongous=0pt plus 1000pt minus 1000pt
\def\caja{\mathsurround=0pt}
\def\eqalign#1{\,\vcenter{\openup2\jot \caja
       \ialign{\strut \hfil$\displaystyle{##}$&$
       \displaystyle{{}##}$\hfil\crcr#1\crcr}}\,}
\newif\ifdtup

\def\1ov4{{1\over 4}}

\def\ddt{\dot{\t}}

\renewcommand{\a}{\alpha}
\renewcommand{\b}{\beta}

\renewcommand{\d}{\delta}
\newcommand{\rmd}{{\rm d}}

\newcommand{\beq}{\begin{equation}}
\newcommand{\eeq}{\end{equation}}

\def\ba{\begin{eqnarray}}
\def\ea{\end{eqnarray}}

\begin{document}
 



\null\vskip-24pt
\hfill KL-TH 00/05
\vskip-10pt
\hfill {\tt hep-th/0005256}
\vskip0.3truecm
\begin{center}
\vskip 3truecm
{\Large\bf
The trace anomaly of the (2,0) tensor
multiplet in background gauge fields}\\ 
\vskip 1.5truecm
{\large\bf
Ruben Manvelyan
 \footnote{On leave of absence from Yerevan Physics Institute, email:{\tt
manvel@physik.uni-kl.de}   
} and Anastasios C. Petkou
   \footnote{email:{\tt
       petkou@physik.uni-kl.de}}  
}\\
\vskip 1truecm
{\it Department of Physics, Theoretical Physics\\
University of Kaiserslautern, Postfach 3049 \\
67653 Kaiserslautern, Germany}\\

\end{center}
\vskip 2truecm
\centerline{\bf Abstract}

We study the trace anomaly of the (2,0) tensor multiplet in $d=6$ in
the presence of a background $SO(5)$ vector field acting as a
source for the $R$-current. Using both a free-field theory
calculation and AdS$_7$/CFT$_6$ correspondence, we find that only one
of the two possible anomaly structures is non-zero and that its
coefficient at 
strong-coupling differs by the well-known overall factor
$4N^3$ from the corresponding weak coupling result. We also discuss the
relevance  
of our result to studies of the $R$-current anomaly in the
(2,0) multiplet.        

\newpage

\section{Some general remarks}

The conjectured AdS/CFT correspondence \cite{mald} provides a rare tool for
studying the strong coupling dynamics of certain gauge theories. Up to
now, the correspondence has found its best application in studies of
the conjectured duality between type 
$IIB$ string theory on AdS$_5\times$S$^5$ describing the dynamics of $N$
coincident $D3$-branes and the large-$N$, large-$g_{YM}^2N$ limit of
${\cal N}=4$ SYM in $d=4$ with gauge group $SU(N)$. This
so-called AdS$_5$/CFT$_4$ correspondence 
is the prototype example for all possible dualities between  
various compactifications of string/$M$-theory and 
gauge field theories.

Of
particular importance in establishing the AdS$_5$/CFT$_4$
correspondence are studies of the trace and the $SO(6)$
$R$-current anomalies in ${\cal N}=4$ SYM$_4$. Such studies have
revealed, among others, 
the field theory 
interpretation of the parameter $N$,  which in the string theory
picture  corresponds to the number of the coincident $D3$-branes,
as being the dimension of the gauge group. Since the operators
in the gauge 
theory transform under the 
adjoint representation of an $SU(N)$ gauge group, the conformal
anomaly in the large-$N$ strong
coupling regime differs from the corresponding weak-coupling value by
an overall $N^2$ factor. Then, as the conformal
anomaly is in the same supermultiplet with the $R$-current anomaly
\cite{freed1},
the large-$N$ strong coupling value of the latter also differs by the
same overall factor $N^2$ from the corresponding weak-coupling value
\cite{freed,chalm,bilal,ManPet1}.   

Recently, {\cite{toine,BFT1,kostas,BFT2,arkady,cappelli} there has been growing
  interest in studying the 
AdS$_{7}$/CFT$_{6}$ correspondence as another example of the
duality between string/$M$-theory and gauge field theory. Explicitly,
this form of the correspondence is conjectured to 
encode  the duality between $M$-theory compactifications on AdS$_7\times$S$^4$
and the maximally supersymmetric (2,0) tensor multiplet in $d=6$. The
former theory describes the low energy limit of $N$ coincident
$M5$-branes. The latter is a mysterious, strongly coupled
six-dimensional CFT without a free coupling parameter. Nevertheless, 
there also exists in $d=6$ a free 
(2,0) tensor multiplet which would, presumably, describe
the weak-coupling regime of the above mysterious theory. 
In studying the AdS$_7$/CFT$_6$ correspondence one naturally
focuses on the conformal and 
$R$-symmetry anomalies of the boundary CFT$_6$ theory. Conformal
anomaly studies  have  
revealed a remarkable property \cite{kostas,BFT2,arkady}: the
Weyl-invariant part of the conformal 
anomaly in the 
strong-coupling regime of the (2,0) multiplet differs by an overall
$4N^3$ factor from the corresponding weak-coupling anomaly, the latter being
calculated using free-fields. The field-theoretic interpretation of
the overall $4N^3$  
factor - yet alone of the $N$ parameter - is far from obvious in this
case and it is conceivably related to
some as yet unknown 
realization of 
gauge invariance in higher dimensions. 

As supersymmetry relates the
trace and the $SO(5)$ $R$-current anomalies (the energy momentum and
the $R$-current are in the same supermultiplet), it is
important that studies of the trace anomaly are compatible with the
well-known result \cite{witten,minasian} for the the $R$-current anomaly of the (2,0)
theory. In this work we present an
explicit calculation of the trace anomaly of the (2,0) tensor
multiplet in the presence of a background $SO(5)$ vector field, both
using a free-field realization and also using 
AdS$_7$/CFT$_6$ correspondence. In this way our calculation yields
respectively the weak- and the strong-coupling results for the trace
anomaly. In the next section we briefly review the structure of the
trace anomaly in $d=6$ in the presence of background vector fields and discuss
its connection to the coefficients of the two- and three-point functions
of $R$-currents. Then we present our calculation. The free-field
calculation is done using Seeley-de Witt coefficients while for the
strong-coupling calculation we rely on AdS$_7$/CFT$_6$
correspondence. We find that only one of the two possible trace
anomaly structures is non-zero. We attribute the result to the maximal
supersymmetry of the (2,0) tensor multiplet. The strong-coupling
result differs from the weak-coupling one by an overall factor
$4N^3$. Finally, we discuss the relevance of our result to the
well-known results for the $R$-current anomaly of the (2,0) tensor
multiplet in $d=6$.

\section{Trace anomaly and $R$-current correlation functions in the
  (2,0) tensor multiplet}

Let $V_{\mu}(x)=V^{A}_{\mu}(x) T^{A}$, $\mu=1,..,d$ be a
general conserved 
current of a $d$-dimensional CFT. In the case when this coincides with
the $R$-current of the (2,0) 
multiplet in $d=6$,  $T^A$ denote the adjoint generators of $SO(5)$. 
When the theory is coupled to a background vector field
$A_{\m}^A(x)$, even in flat spacetime the trace of the energy momentum
tensor acquires an 
(external) anomaly which up to total derivative terms can be written as 
\cite{ManPet1} 
\ba
\langle T_{\m}^{\,\,\m}(x)\rangle =
\alpha_{V}\,F^{A,\,\n}_{\m}F^{B,\,\l}_{\n}F^{C,\,\m}_{\l}f^{ABC}
+  
\beta_{V}\,\nabla^{\m}F^{A}_{\m\n}\,\nabla^{\l}F^{A,\,\n}_{\l}\,,
\label{t}
\ea
where $F_{\m\n}^A(x)$ is the standard field strength of
$A_{\m}^A(x)$. Notice in (\ref{t}) the presence of two different structures in
in contrast to the $d=4$ case where only one structure
appears. An important property of the parameters $\a_V$, $\b_V$ in
(\ref{t}) is that they are intimately connected to the parameters
appearing in the two- and three-point functions of the $R$-current
$V_{\m}^A(x)$.   Such a
connection follows from the observation that the external trace
anomaly is tied to the 
short distance singularities of renormalized $n$-point functions. In
the case of interest here, assuming a coupling to the background
vector field of the form $
\int\rmd^{d}x\,\sqrt{g}\,g^{\m\n}A^{A}_{\m}(x)V^{A}_{\n}(x)$
we can follow \cite{petsken} and write
the general 
renormalization group equation as 
\ba
&&\hspace{-0.8cm} \sum_{k=1}^{\infty}\frac{1}{k!}
\int\rmd^{6}x_{1}\sqrt{g}g^{\m_{1}\n_{1}}  
\,..\,\rmd^{6}x_{k}\sqrt{g}g^{\m_{k}\n_{k}}
A^{A_1}_{\m_{1}}(x_{1})..
A^{A_k}_{\m_{k}}(x_{k})\,\m\frac{\partial}{\partial\mu}
\langle V^{A_1}_{\n_{1}}(x_{1})..V^{A_k}_{\n_{k}}(x_{k})
\rangle_{R}= \nonumber\\
&&\hspace{1.2cm}=\int\rmd^{6}x\sqrt{g}g^{\m\n}(x)\langle 
T_{\m\n}(x)\rangle\,.\label{rg} 
\ea 
The
subscript $R$ in the first line of (\ref{rg}) denotes the renormalized
$n$-point functions which 
depend on the arbitrary mass parameter $\m$. Taking suitable
functional derivatives of (\ref{rg}) with respect to $A_{\m}^{A}(x)$
we can, in principle, connect the parameters which appear in $n$-point
functions of $J_{\mu}^{A}(x)$ with the possible terms 
in the trace anomaly. Then, the  importance of (\ref{rg}) is based on
the fact that in a CFT the two- and three-point functions of conserved
currents are determined for general $d$  up to a number of constant
parameters. Specifically, for the case of the conserved current
$V_{\m}^A(x)$ one has \cite{ospet}
\ba
\<V^{A}_{\m}(x)V^{B}_{\n}(y)\> &=& \d^{AB}\frac{{\cal
C}^{(d)}_{V}}{[(x-y)^2]^{(d-1)}}I_{\m\n}(x-y), \,\,\,\,I_{\m\n}(x)=\d_{\m\n}
-2\frac{x_{\m}x_{\n}}{x^{2}}\,, \label{corr1}\\  
\<V^{A}_{\m}(x)V^{B}_{\n}(y)V^{C}_{\l}(z)\> &=&
\frac{f^{ABC}}{(x-y)^{d-2}(x-z)^{d-2}(y-z)^{d}} \nonumber\\  
&&\times \, I_{\n\s}(x-y)\,I_{\l\r}(x-z)\,
I_{\s\b}(X)\,t_{\m\b\r}(X)\,,\label{corr2}\\  
X_{\m}&= &\frac{(x-y)_{\m}}{(x-y)^2} -
\frac{(x-z)_{\m}}{(x-z)^2} \,,\nonumber\\ 
t_{\m\n\l}(X) &=& {\cal A}^{(d)}\frac{X_{\m}X_{\n}X_{\l}}{X^{2}} + {\cal
B}^{(d)}\left( X_{\m}\d_{\n\l} + X_{\n}\d_{\m\l} -
X_{\l}\d_{\m\n}\right)\,,\nonumber 
\ea
Therefore, had one been able to provide the explicit relation between
the parameters ${\cal C}^{(6)}_{V}$, ${\cal A}^{(6)}$, ${\cal
B}^{(6)}$ and $\a_V$, $\b_V$, the calculation of the latter two would 
reduce to a calculation of the former three. In such a case, one
could utilize the well-known AdS$_7$/CFT$_6$ calculations for the two-
and three-point functions of conserved currents \cite{freed} to calculated the
trace anomaly.  

Using differential renormalization arguments, the parameter $\b_V$ was evaluated in terms of
${\cal C}^{(6)}_V$ as \cite{ManPet1}
\ba
\beta_{V} = \frac{{\cal C}^{(6)}_{V}\pi^3}{960}\,.
\label{betafinal}
\ea 
The corresponding result for $\a_V$ is still missing, however on
general grounds one expects it to be a linear combination of ${\cal
  A}^{(6)}$ and $ {\cal B}^{(6)}$.  

The above show that both the weak- and the strong-coupling values
of the trace anomaly parameters $\a_V$ and $\b_V$ for the (2,0) tensor
multiplet can be obtain from
the corresponding values of ${\cal C}^{(6)}_{V}$, ${\cal
  A}^{(6)}$ and ${\cal B}^{(6)}$. The weak-coupling values of the
latter three parameters can be found using a free-field realization
for the theory as \cite{ManPet1}
\ba
{\cal C}^{(6)}_{V,free}&=&\frac{5}{\pi^6},\qquad {\cal B}^{(6)}_{free}
=\frac{9}{2\pi^9},\qquad  {\cal A}^{(6)}_{free}=\frac{3}{\pi^9}\,.
\label{2.0num}
\ea
The strong-coupling values can be found using
AdS$_7$/CFT$_6$ 
correspondence which requires the consideration of maximal supergravity
on AdS$_7$.\footnote{We consider
the Euclidean 
version  of 
AdS$_{d+1}$  space where  $\rmd \hat{x}^{i}\rmd
\hat{x}_{i}=\frac{1}{x_{0}^{2}} 
(\rmd x_{0}\rmd 
x_{0}+\rmd x^{\m}\rmd x^{\m})$, with $\m=1,..,d.$ and
$\hat{x}_{i}=(x_{0},x_{\m})$. 
The boundary of this space is isomorphic
to {\bf{S}}$^{\rm d}$ since it consists of
{\bf{R}}$^{\rm d}$ at $x_{0}=0$ and a single point at $x_{0}=\infty$.}
The relevant part of the supergravity Lagrangian is  
\ba
 {\cal L}=-\frac{1}{4g^{2}_{SG_{7}}}\int \rmd^{7}\hat{x}
\,\sqrt{g}\,F^{A}_{ij}(\hat{x} )F^{A,ij}(\hat{x})\,, \quad i,j=0,..,6\,,
\label{lag}  
\ea
from where  one obtains \cite{freed,chalm}
\beq
{\cal C}^{(6)}_{V}=\frac{120}{\pi^{3}g^{2}_{SG_{7}}},\,\,\,
{\cal A}^{(6)}= 
\frac{72}{\pi^{6}g^{2}_{SG_{7}}},\,\,\, {\cal B}^{(6)}=
\frac{108}{\pi^{6}g^{2}_{SG_{7}}}\,.\label{ads7num}
\eeq

The important remaining piece of information is the value of
$1/g_{SG_{7}}^2$ . This can be read from the general results for
$d+1=4,5,7$, given in \cite{ManPet1}. One considers the equations of
motion of gauged 
supergravity in 4, 5 and 7 dimensions correspondingly in the presence of
non trivial scalar fields in the coset space $SL(n,{\bf
R})/SO(n)$. The $SO(n)$ group corresponds to the $R$-symmetry group of
the boundary CFT$_d$. 
The result is
\beq
\label{new2}
\frac{1}{g_{SG_{d+1}}^2} =\frac{n(n-2)}{4d(d-1)} \frac{1}{2\kappa_{d+1}^2}=
\frac{1}{2\kappa_{d+1}^2}\frac{2}{(d-2)^2}\,,
\eeq
where we have used the relation $n=4\frac{d-1}{d-2}$ \cite{gleb2}.
Then, from (\ref{2.0num}) and (\ref{ads7num}) we obtain \footnote{It
  is important to point out 
  that the general result (\ref{new2}) is compatible with all known
  calculations of two- and three-point functions in $d=3,4,6$ (see
  e.g. \cite{gleb1}). In   
  particular, for $d=3$ we obtain $\frac{{\cal C}^{(3)}_{V}}{{\cal C}^{(3)}_{V,free}}=\frac{{\cal
B}^{(3)}}{{\cal B}^{(3)}_{free}}=\frac{{\cal
A}^{(3)}}{{\cal A}^{(3)}_{free}}=\frac{N^{\frac{3}{2}}4\sqrt{2}}{3\pi}
$ which shows that the ratio between the strong- and weak-coupling
values for the two- and three-point functions in $d=3$ is
$\frac{N^{\frac{3}{2}}4\sqrt{2}}{3\pi}$. This irrational overall
factor coincides with the one found in \cite{BFT1} in studies of the
two- and three-point functions of the energy momentum tensor.}
\ba
\frac{{\cal C}^{(6)}_{V}}{{\cal C}^{(6)}_{V,free}}&=&\frac{{\cal
B}^{(6)}}{{\cal B}^{(6)}_{free}}=\frac{{\cal
A}^{(6)}}{{\cal A}^{(6)}_{free}}=4N^3 \,.
\label{n3}
\ea 
Then, from (\ref{betafinal}) and (\ref{n3}) we obtain
\ba
\beta_{V} = 4N^3\b_{V,free}=\frac{N^3}{192\pi^3}\,,
\label{betafinal2}
\ea 
Our result shows that the strong-coupling value of the trace anomaly
parameter $\b_V$  differs from its weak-coupling value by an overall
$4N^3$ factor.

\section{The free-field result}

In the absence of a result such as (\ref{betafinal}) for the trace
anomaly parameter $\a_V$, we have to rely on some other method
for calculating the trace anomaly (\ref{t}) both in the weak- and also
in the strong-coupling regimes of the (2,0) tensor multiplet. In each
case, agreement with the result (\ref{betafinal2}) would be a strong
test for our calculation.

The weak-coupling values of both coefficients $\alpha_{V}$ and $\beta_{V}$ 
can be calculated by the method of Seeley-De Witt coefficients using a
free-field realization of the (2,0) tensor 
multiplet.
Following \cite{BFT2} we can evaluate the anomaly in $d=6$ from the 
Seeley-De Witt coefficient $b_{6}$ for the general second order
Laplace operator 
\ba
\label{lapl}
\D=-\nabla^2 -E,
\ea
where $\nabla_{\m}$ is a covariant derivative with normal bundle connection
$[\nabla_{\m},\nabla_{\n}]=F_{\m\n}$ and $E$ is a matrix endomorphism.
The general formula for $b_{6}$  is (see \cite{BFT2} and references therein)
\ba
b_6 &=& { 1\over (4\pi)^3}
(\a_6 +{1\over 6} \alpha_2^3 + \alpha_2 \alpha_4)\,,\label{b6}\\
\a_2 &=& E\,,\label{a22}\\
\a_4 &=& {1\over 6} \nabla^2 E + {1\over 12} F_{\m\n}^2 \,,\label{a4}\\
\a_6 &=& {2\over 6!} \left[ 8 (\nabla_\a F_{\m\n})^2 +2 (\nabla^\a F_{\a\m})^2
+12 F_{\a\b}\nabla^2 F^{\a\b} - 12 F_\a{}^\m F_\m{}^\b
F_\b{}^\a\right.\label{a6}\\  
&&\left. + 6 \nabla^4 E  + 30 (\nabla_\a E)^2 \right]\,.\nonumber
\ea
We have to evaluate $b_6$ for fermions and bosons taking into 
account the following identities
\ba
\nabla_\a F_{\m\n}\nabla^\a F^{\m\n}&=& 2\nabla^\m F_{\m\a}\nabla^\n F_\n{}^\a
-2 F_\a{}^\m F_\m{}^\b F_\b{}^\a +\nabla_\a J_{1}^\a \,,\label{id}\\
J_{1}^\a &=& F_{\n\m}\nabla^\m F^{\n\a} - F_\n{}^\a\nabla_\m F^{\n\m}\,.
\ea
The last total derivative term in (\ref{id}) is not important here as
it can be cancelled by adding local counterterms \cite{anom},
therefore we can drop it in our calculation. 

For scalar bosons we have
\ba
F_{\m\n} = F_{\m\n}^{A}T^A ,&& E=0\label{s1}\,,\quad
Tr\left( F_{\m\n}F^{\m\n}\right) = -C_\fii F_{\m\n}^A F^{A,\m\n} \,
,\label{s10}\\ 
Tr\left(F_\a{}^\m F_\m{}^\b F_\b{}^\a\right)&=&
-{1\over 2}C_{\fii} f^{ABC}F_{\a}^{A,\m} F_{\m}^{B,\b}
F_{\b}^{C,\a}\,,\label{s2}\\ 
b^{s}_{6}&=&\frac{C_{\fii}\left(\nabla^{\a}F_{\a\m}^{A}\right)^2}{(4\pi)^3
  60} + 
\frac{C_{\fii}f^{ABC}F_{\a}^{A,\m} F_{\m}^{B,\b} F_{\b}^{C,\a}}{(4\pi)^3
  180}\,. \label{s3}
\ea
The corresponding spinor contribution is
\ba
F_{\m\n} = F_{\m\n}^{A}T^A {\bf I}_{\j},&&  E= {1\over 2}F_{\m\n}^A T^A  \gamma^{\m\n}\,,\quad
 Tr\left( F_{\m\n}F^{\m\n}\right) = -C_\j ({\rm Tr}{\bf I}_{\j})
 F_{\m\n}^AF^{A,\m\n} \,, \label{f1}\\
\qquad  Tr\left(F_\a{}^\m F_\m{}^\b F_\b{}^\a\right)&=&
-{1\over 2}C_{\j }({\rm Tr}{\bf I}_{\j}) f^{ABC}F_{\a}^{A,\m}
F_{\m}^{B,\b} F_{\b}^{C,\a}\,, \label{f2}\\   
b^{f}_{6}&=& -\frac{C_{\j}({\rm Tr}{\bf
    I}_{\j})\left(\nabla^{\a}F_{\a\m}^{A}\right)^2}{(4\pi)^3 15} + 
\frac{C_{\j}({\rm Tr}{\bf I}_{\j})f^{ABC}F_{\a}^{A,\m} F_{\m}^{B,\b}
  F_{\b}^{C,\a}}{(4\pi)^3 180}\,.\label{f3} 
\ea
From (\ref{s3}) and (\ref{f3}) we can calculate the free-field
result for the total trace 
anomaly of the $d=6$, $(2,0)$ tensor multiplet in the the presence of
external vector fields and up to total derivatives terms we obtain
\ba 
\label{an6}
\langle T_\m{}^\m(x)\rangle &=&
b^{s}_6(x)-b^{f}_6(x)\label{a1}\nonumber \\
&=&\frac{\left(\nabla^{\a}F_{\a\m}^{A}\right)^2}{(4\pi)^3
  15}\left({C_{\fii}\over 4} + C_{\j }({\rm Tr{\bf I}_{\j}})\right)  + 
\frac{f^{ABC}F_{\a}^{A,\m} F_{\m}^{B,\b} F_{\b}^{C,\a}}{(4\pi)^3
  180}\Bigl(C_{\fii} - C_{\j }({\rm Tr}{\bf I}_{\j})\Bigl)
\label{a2}\nonumber \\
&=& \frac{{\cal
    C}^{(6)}_{V,free}\pi^3}{960}\left(\nabla^{\a}F_{\a\m}^{A}\right)^2
+ 0\cdot f^{ABC}F_{\a}^{A,\m} F_{\m}^{B,\b} F_{\b}^{C,\a}\,.\label{a3}  
\ea
The get from the second line to the third in (\ref{a3}) we used the
general expression 
for  ${\cal C}^{(6)}_{V,free}$ \cite{ospet} and also the following selection
rule obtained in 
\cite{ManPet1}
\ba 
C_{\fii} = C_{\j}({\rm Tr}{\bf I}_{\j})\,,
\label{selrule}
\ea 
for the free-field realization of the $(2,0)$ tensor multiplet.
The value of $\b_V$ read off from (\ref{a3}) coincides up to an
overall $4N^3$ factor with
(\ref{betafinal2}), which is a consistency test for our
calculation. We also obtain  
$\alpha_V = 0$ as a direct result of the selection rule (\ref{selrule}).

\section{The strong-coupling result from AdS$_7$/CFT$_6$ correspondence}

To calculate the anomaly coefficients in the strong coupling limit
we use  the  $AdS$ action (\ref{lag}) and follow the methods developed
in \cite{kostas,kostas2}. For that, we have to consider the on-shell
dependence of (\ref{lag}) 
on the boundary value of the gauge field and then extract the
logarithmic divergence. 
This is achieved by solving the equations of motion for the Yang-Mills field
$A_{i}(\hat{x}) \equiv A^{A}_{i}T^{A}=(A_{0}(x_0,
x_{\m}),A_{\m}(x_0, x_{\m}))$ 
in $AdS$. These equations are significantly simplified by choosing to
work on the gauge $A_{0}=0$, which is a natural gauge condition
preserving gauge invariance on the boundary. In this gauge the equations of
motions following from (\ref{lag}) are 
\ba
\hat{\nabla}_{j}F^{ji}&\equiv&{1\over
  \sqrt{g}}\partial_{j}\left(\sqrt{g}F^{ji}\right) +
\left[A_{j},F^{ji}\right]\,\,=\,\,0\,,\label{em1}\\ 
\hat{\nabla}_{\m}F_{\m x}&=&0\,,\quad x_0^{d+1}\partial_{0}\left(
  x_0^{-d-1+4}F_{x\m}\right) +
x_0^4\hat{\nabla}_{\n}F_{\n\m}\,\,=\,\,0\,. \label{em2}
\ea
Then, following \cite{kostas2} (see also \cite{marika}),  we expand the
vector fields around 
the conformal boundary in a power series as  
\ba
A_{\m}(x_0, x_{\m})&=&\sum_{k=0}^{\infty}x_0^k A^{(k)}_{\m}(x_{\m})\,\,\,,\quad
F_{0\m}=\partial_{0}A_{\m}=\sum_{k=1}^{\infty}
kx_0^{k-1}A^{(k)}_{\m}(x_{\m})\label{A}\,,\\ 
F_{\m\n}(x_0, x_{\m})&=&\sum_{k=0}^{\infty}x_0
^kF^{(k)}_{\m\n}(x_{\m})=\sum_{k=0}^{\infty}x_0^k
\left(\partial_{\m}A^{(k)}_{\n}-\partial_{\m}A^{(k)}_{\n}+\sum_{l=0}^{k}
\left[A^{(l)}_{\m},A^{(k-l)}_{\n}\right]\right)\,.\label{F}
\ea
The boundary value of the vector field is
$A^{(0)}_{\m}(x)$, consequently in order to be able to extract the
logarithmic singularity of the action (\ref{lag}) we have in the
following to introduce
a suitable IR regulator in the $x_0$-integration. Here we just mention
that a simple  
expansion such as (\ref{A}) implemented with a suitable IR
regularization of the $x_0$-integration can be easily shown to give
the correct conformal anomaly for massless scalars in any dimension. 
Substituting (\ref{A}) and (\ref{F}) into (\ref{em2}) we obtain
\ba
\sum_{k=1}^{\infty}(2+k-d)kx_0^{k+2}A_{\m}^{(k)}&=&-
\sum_{k=1}^{\infty}x_0^{k+3}\nabla_{\n}F_{\n\m}^{(k-1)} 
- \sum_{k=1}^{\infty}x_0^{k+4}\sum_{l=0}^{k}\left[A_{\n}^{(l)},
  F_{\n\m}^{(k-l)}\right]\,,\label{gensol}\\ 
\sum_{k=1}^{\infty}kx_0^{k-1}\nabla_{\m}A_{\m}^{(k)}&=&-
\sum_{k=2}^{\infty} x_0^{k-1}\sum_{l=1}^{k-1}l\left[A_{\m}^{(k-l)},
  A_{\m}^{(l)}\right]\,,\label{gensol1} 
\ea
where $\nabla_{\m}=\partial_{\m}+\left[A^{(0)}_{\m},...\right]$.
The equations (\ref{gensol}), (\ref{gensol1}) can be recursively
solved for all $k$. Here we consider only the relevant to us cases
$k=1,2,...$ when we obtain 
the following set of solutions   
\ba
A_{\m}^{(1)}&=&0\,,\label{A1}\\
A_{\m}^{(2)}&=&-{1\over 2(4-d)}\nabla_{\n}F_{\n\m}^{(0)} \,\,\,\,{\buildrel
  {(d=6)}\over =}\,\,\,\, {1 \over
  4}\nabla_{\n}F_{\n\m}{(0)}\,,\,\,\nabla_{\m}A_{\m}^{(2)}=0 \,,\label{A2}
\\A_{\m}^{(3)}&=&0\,.\label{A3}
\ea
Plugging in (\ref{A1})-(\ref{A3}) into (\ref{lag}) we can extract the
logarithmic term as 
\ba
-{1\over {4g^{2}_{SG_7}}}\int_{\e}^{\infty} {dx_0 \over x_0} \int d^6x\left(
  8A_{\m}^{(2)}A_{\m}^{(2)} +  
4F^{(0)}_{\m\n}\nabla_{\m}A^{(2)}_{\n}\right) =\ln
\epsilon{1\over{8g^{2}_{SG_7}}}\int d^6x\left(\nabla^\m
  F^{(0)}_{\m\n}\right)^2\,.\label{lg} 
\ea
In (\ref{lg}) we used the standard procedure to obtain the
regularized generating functional for the boundary CFT$_6$ by
considering the $x_0$-coordinate as an IR regulator in the 
bulk ($x_0\in(\infty,\e),\e\rightarrow 0$) which corresponds to an UV
regulator in the boundary $\m=1/\e$ dropping at the same time all
bulk UV divergences. Then, from (\ref{ads7num}) we finally get by
virtue of (\ref{n3}) 
\ba
\langle T_{\m}^{\,\,\m}(x)\rangle_{AdS_7} =
0\cdot F^{A,\n}_{\m}F^{B,\l}_{\n}F^{C,\m}_{\l}f^{ABC}
+  
\frac{{\cal
 C}^{(6)}_{V}\pi^3}{960}\,
\left(\nabla^{\m}F^{A}_{\m\n}\right)^2\,.   
\label{adsb}
\ea
 
\section{Discussion}

Our results (\ref{a3}) and (\ref{adsb}) for the trace anomaly of the
(2,0) multiplet in the presence of external gauge fields were obtained
in the absence of external gravity. They are
characterized both by the manifestation of the overall $4N^3$ as one
goes from the weak-coupling (free-fields) to the strong-coupling
regime and also by the vanishing of the one of the two possible
structures (namely $\a_V=0$). The latter fact can be attributed, as
seen from 
(\ref{a3}), to the selection rule (\ref{selrule}) which in turn may be viewed
as a manifestation of maximal supersymmetry both in the supergravity
and also in the boundary CFT$_d$.

Moreover, the results above are connected to the
well-known results for the $R$-current anomaly \cite{freed1}. The structure of the
supersymmetry algebra in $d=6$ is, however,  quite involved
and an explicit 
relation between the trace and the $R$-symmetry anomalies has not
appeared in the literature as yet. Nevertheless, a separate discussion
of the known results for the 
trace and $R$-current anomalies of the (2,0) multiplet might be
useful.

In the
absence of an external gravitational background the $R$-current
anomaly is given by the following 8-form
\ba
I_8^{free}(F)&=&\frac{1}{3\cdot2^4}\left[p_2(F)
  +\frac14p_1(F)^2\right] \,,\label{I8free}\\
p_1(F) & = & \frac12{\rm
  tr}\tilde{F}^2\,,\,\,\,\,p_2(F)=-\frac14\left[{\rm tr}\tilde{F}^4
  -\frac12 {\rm tr}\tilde{F}^2 \wedge {\rm tr}\tilde{F}^2
\right]\,,\,\,\,\, \tilde{F}=\frac{\rm i}{2\pi}F\,.\label{p1p2}
\ea
(\ref{I8free}) gives the anomaly ${\cal I}_6$ in the six-dimensional
theory via the 
descent equations $d(\d {\cal I}_6) =\d I_7$, $I_8 =\d I_7$. Notice
that, at first sight, the descent formalism seems to give two
possible linearly independent  structures for the six-dimensional
anomaly. This has to be compared with our trace anomaly results (\ref{a2}) and
(\ref{adsb}) which involve only one structure.  

The $R$-symmetry anomaly of the strongly-coupled (2,0) multiplet has been also
evaluated requiring the cancellation of the total anomaly of $N$
$M5$-branes when one takes into account the inflow anomaly
\cite{witten}. The result 
in the absence of external gravity is \cite{minasian}
\beq
I_8^{(2,0)}(F) = \frac{1}{3\cdot 2^4}\left[(2N^3-N) p_2(F)
  +\frac{N}{4}p_1(F)^2\right]\,.\label{I8}
\eeq
Now, if AdS$_7$/CFT$_6$ correspondence is valid, one should be able to
recover the large-$N$ limit of the latter result by considering the
maximally supersymmetric $N=2$ gauged SUGRA in $d=7$. Namely, one
should find a result which lifted to 8-dimensions should read
\beq
I_8^{(AdS/CFT)}(F) = \frac{2N^3}{3\cdot 2^4}p_2(F)\,.\label{I8N}
 \eeq
One then observes that (\ref{I8free}) and (\ref{I8N}) seem
to imply that the $R$-symmetry anomaly 
has different structure in the weak (free-fields) and the 
strong-coupling regimes. This in turn implies some kind of
renormalization of the 
$R$-symmetry anomaly as one goes from the weak to the strong coupling
regimes and it is reminiscent to the 
corresponding result for the 
trace anomaly of the (2,0) multiplet in an external gravitation
background in which at least the Euler density term seems to be also
renormalized \cite{BFT2,arkady}. Moreover, by virtue of supersymmetry
(\ref{I8free}) and 
(\ref{I8N}) seem to indicate that the 
corresponding result for the trace anomaly in the presence of external
vector fields, but in the absence of external gravity, would also be
renormalized as one goes from the weak to the strong-coupling
regimes. Such a conclusion appears to be, at first sight, incompatible
with our result (\ref{a3}) and (\ref{adsb}) i.e. that the structure 
of the trace anomaly is the same in both the weak and the strong
coupling regimes. It is conceivable that  a better understanding
of supersymmetry in $d=6$ might resolve this apparent puzzle.

\section*{Acknowledgments}
We wish to thank G. Arutyunov and R. Minasian for very enlightening
discussions and also S. Frolov and especially A. Tseytlin for very
useful correspondence. 
We would also like to thank the organizers of the Workshop ``Duality,
Strings and $M$-Theory'' in E.S.I., Vienna, where this work was
initiated, this for their worm hospitality and
financial support.  The work of R. M. was partially supported by
Alexander von Humboldt Foundation and by the INTAS grant
No:99-590. The work of A. C. P. was supported by Alexander von
Humboldt Foundation.

\end{document}